\documentclass[aps,prd,twocolumn,floatfix,nofootinbib,superscriptaddress,showpacs,amssymb]{revtex4}

\usepackage{bm}
\usepackage{graphicx}

\def\beqa{\begin{eqnarray}}
\def\eeqa{\end{eqnarray}}
\def\beq{\begin{equation}}
\def\eeq{\end{equation}}
\def\p{\partial}
\def\vp{\varphi}

\def\ulap{\underline\Delta}
\def\unab{\overline\nabla}

\newcommand{\w}[1]{\bm{#1}}

\begin{document}

\title{Last orbits of binary strange quark stars}

\author{Fran\c cois Limousin}
\email[]{Francois.Limousin@obspm.fr}
\affiliation{Laboratoire de l'Univers et de ses Th\'eories,
UMR 8102 du C.N.R.S., Observatoire de Paris, Universit\'e Paris 7, F-92195 Meudon Cedex, France}
\author{Dorota Gondek-Rosi\'nska}
\email[]{Dorota.Gondek@obspm.fr}
\affiliation{Laboratoire de l'Univers et de ses Th\'eories,
UMR 8102 du C.N.R.S., Observatoire de Paris, Universit\'e Paris 7, F-92195 Meudon Cedex, France}
\affiliation{Nicolaus Copernicus Astronomical Center, Bartycka 18,
00-716 Warszawa, Poland}
\affiliation{Departament de Fisica Aplicada, Universitat d'Alacant, Apartat de correus 99, 03080 Alacant, Spain} 
\affiliation{Institute of Astronomy, University of Zielona G\'ora, Lubuska 2, 65-265, Zielona G\'ora, Poland}
\author{Eric Gourgoulhon}
\email[]{Eric.Gourgoulhon@obspm.fr}

\affiliation{Laboratoire de l'Univers et de ses Th\'eories,
UMR 8102 du C.N.R.S., Observatoire de Paris, Universit\'e Paris 7, F-92195 Meudon Cedex, France}


\begin{abstract}
We present the first relativistic calculations of the final phase of
inspiral of a binary system consisting of two stars built
predominantely of strange quark matter (strange quark stars).  We study
the precoalescing stage within the Isenberg-Wilson-Mathews approximation of
general relativity using a multidomain spectral method. 
A hydrodynamical treatment is performed under the assumption that the
flow is either rigidly rotating or irrotational, taking into account
the finite density at the stellar surface --- a distinctive feature
with respect to the neutron star case. The gravitational-radiation 
driven evolution of the binary system is approximated by a sequence of 
quasi-equilibrium configurations at fixed baryon number and decreasing
separation. We find that the innermost stable circular orbit (ISCO)
is given by an orbital instability both for synchronized and
irrotational systems. This constrasts with neutron stars for which 
the ISCO is given by the mass-shedding limit in the irrotational case. 
The gravitational wave frequency at the ISCO, which marks the end of
the inspiral phase, is found to be $\sim 1400$~Hz for two irrotational
$1.35\, M_\odot$
strange stars and for the MIT bag model of strange matter
with massless quarks and a bag constant $B=60\ {\rm MeV\,
fm^{-3}}$. Detailed comparisons with binary neutrons star models, 
as well as with third order Post-Newtonian point-mass binaries are
given. 
\end{abstract}


\maketitle

\section{Introduction}

One of the most important prediction of general
relativity is gravitational radiation.  Coalescing neutron star
binaries are considered among the strongest and most likely sources of
gravitational waves to be seen by VIRGO/LIGO interferometers
\cite{Kalogera04, Belczynski02}.  Due to the emission of
gravitational radiation, binary neutron stars decrease their orbital
separation and finally merge. Gravitational waves emitted during the
last few orbits of inspiral could yield important informations about
the equation of state (EOS) of dense matter 
\cite{Faber02,TanigG03,Oechslin04,Bejger04}. With accurate templates
of gravitational waves from coalescing binary compact stars, it may be
possible to extract information about physics of neutron stars from
signals observed by the interferometers and to solve one of the
central but also most complex problem of physics --- the problem of the
absolute ground state of matter at high densities. It is still an open
question whether the core of a neutron star consists mainly of
superfluid neutrons or exotic matter like strange quark matter, pions
or kaons condensates (see e.g. Ref.~\cite{Haens03} for a recent 
review). The possibility of the existence of quark
matter dates back to the early seventies. Bodmer \cite{Bodmer71}
remarked that matter consisting of deconfined up, down and strange
quarks could be the absolute ground state of matter at zero pressure
and temperature.  If this is true then objects made of such matter,
the so-called {\em strange stars}, could exist
\cite{Witten84,HaensZS86,AlcocFO86}. 
Strange quark stars are currently considered as a
possible alternative to neutron stars as compact objects (see
e.g. \cite{Weber04, Madsen99, Gondek03} and references
therein).

The evolution of a binary system of compact objects is entirely driven
by gravitational radiation and can be roughly divided into three
phases : point-like inspiral, hydrodynamical inspiral and merger. The
first phase corresponds to large orbital separation (much larger than
the neutron star radius) and can be treated analytically
using the post-Newtonian (PN) approximation to general relativity
(see Ref.~\cite{Blanc02b} for a review). In the second phase the orbital separation
becomes only a few times larger than the radius of the star, so
the effects of tidal deformation, finite size and hydrodynamics
play an important role.  In this phase, since the shrinking
time of the orbital radius due to the emission of gravitational waves
is still larger than the orbital period, it is possible to approximate
the state as quasi-equilibrium
\cite{BaumgCSST97,BonazGM99}. 
The final phase of the evolution is the merger of the two objects,
which occur dynamically 
\cite{ShibaU00,ShibaU01,ShibaTU03,OoharN99}. 
Note that quasi-equilibrium computations from the second phase 
provide valuable initial data for the merger 
\cite{ShibaU00,ShibaTU03,Oechslin04,FaberGR04}. 

Almost all studies of the final phase of the inspiral 
of close binary neutron star systems employ a simplified EOS of
dense matter, namely a polytropic EOS 
\cite{BonazGM99,MarroMW99,UryuE00,UryuSE00,GourGTMB01,Faber02, 
TanigG02b,TanigG03,FaberGR04,MarronDSB04}. There are only two
exceptions: (i) Oechslin et al. have used a pure nuclear matter EOS, 
based on a relativistic mean field model and a `hybrid' EOS 
with a phase transition to quark matter at high density \cite{Oechslin04};
(ii) Bejger et al. have computed quasi-equilibrium sequences based
on three nuclear matter EOS \cite{Bejger04}.
In this article we present results on the hydrodynamical phase of 
inspiraling binary strange
quark stars described by MIT bag model. The calculations are performed
in the framework of {\em Isenberg-Wilson-Mathews} approximation to
general relativity (see Ref. \cite{BaumgS03} for a review). We
consider binary systems consisting of two identical stars. We choose
the gravitational mass of each star to be $1.35 \, M_\odot$ in infinite
separation in order to be consistent with recent population synthesis
calculations \cite{BulikGB04} and with the current set of
well-measured neutron star masses in relativistic binary radio
pulsars \cite{Lorim01,Burga03}. 
We compare the evolution of a strange star binary system
with a neutron star binary in order to find any characteristic
features in the gravitational waveform that will help to distinguish between 
strange stars
and neutron stars. We consider two limiting cases of velocity flow in stellar
interior: the irrotational and the synchronized case in order to
exhibit the differences between these two extreme states. The
irrotational case is more realistic since the viscosity of neutron
star matter (or strange star matter) is far too low to ensure
synchronization during the late stage of the inspiral
\cite{BildsC92,Kocha92}. 
Due to the finite density at the surface of bare strange stars, 
we had to introduce a treatment of the boundary condition for the
velocity potential (in the irrotational case) different from that
of neutron stars, where the density vanishes at the stellar surface.

The paper is organized in the following way: Sec. II is a brief
summary of the assumptions upon which this work is based, Sec. III is
devoted to the description of the EOS used to describe strange stars
and neutron stars. In Sec. IV we briefly describe the basic equations
for quasi-equilibrium and derive the boundary condition required for
solving the fluid equation of irrotational flow with finite surface
density, which is relevant for strange stars. In Sec. V we present the
numerical results for corotating and irrotational strange stars
binaries and compare their quasistationary evolution with that of
neutron stars, as well as with that of post-Newtonian point-masses.
Section VI contains the final discussion.  Throughout the paper, we
use geometrized units, for which $G=c=1$, where $G$ and $c$ denote the
gravitational constant and speed of light respectively.

\section{Assumptions}
The first assumption regards the matter stress-energy tensor
$\w{T}$, which we
assume to have the {\bf perfect fluid} form: 
\beq 
    \w{T} = (e+p)\w{u}\otimes\w{u} +p \, \w{g}, 
\eeq 
where $e$, $p$, $\w{u}$ and $\w{g}$ are respectively the fluid proper energy 
density, the fluid pressure, the fluid 4-velocity, and the spacetime metric. 
This is a very good approximation for neutron star matter or strange star
matter.

The last orbits of inspiraling binary compact stars can be studied in
the {\bf quasi-equilibrium} approximation. Under this assumption the
evolution of a system is approximated by a sequence of exactly
circular orbits. This assumption results from the fact that the time
evolution of an orbit is still much larger than the orbital period and that 
the gravitational radiation circularizes an orbit of a binary system.
This implies a continuous spacetime symmetry, called  {\em helical
symmetry} \cite{BonazGM97,FriedUS02}  represented by the Killing
vector: \beq \w{\ell} = {\p \over \p t} +\Omega {\p \over \p \vp}, \eeq
where $\Omega$ is the orbital angular velocity and $\p/\p t$ and
$\p/\p \vp$ are the natural frame vectors associated with the time
coordinate $t$ and the azimuthal coordinate $\vp$ of an asymptotic
inertial observer.

One can then introduce the {\em shift vector} $\w{B}$ of co-orbiting
coordinates by means of the orthogonal decomposition of $\w {\ell}$ with
respect to the $\Sigma_t$ foliation of the standard 3+1 formalism:
\beq \label{e:helicoidal_n} \w {\ell} = N \, \w{n} - \w{B} , \eeq
where $\w{n}$ is the unit future directed vector normal to
$\Sigma_t$, $N$ is called the {\em lapse function} and
$\w{n}\cdot\w{B}=0$.

We also assume that the spatial part of the metric 
(i.e. the metric induced by $\w{g}$ on each hypersurface $\Sigma_t$)
is conformally flat,
which corresponds to the {\bf Isenberg-Wilson-Mathews} (IWM)
approximation to general relativity \cite{Isenb78,IsenbN80,WilsoM89}
(see Ref.  \cite{FriedUS02} for a discussion). Thanks to this
approximation we have to solve only five of the ten Einstein
equations. In the IWM  approximation, the spacetime metric takes the
form: \beq ds^2 =-(N^2 -B_i B^i) dt^2 -2B_i \, dt \, dx^i +A^2 f_{ij}
dx^i dx^j,
  \label{eq:metric}
\eeq 
where $A$ is some conformal factor,
$f_{ij}$ the flat spatial metric and Latin indices run in $\{1,2,3\}$
(spatial indices).  
The comparison between the IWM results presented here and the
non-conformally flat ones will be performed in a future article
\cite{UryuL04}. 

The fourth assumption concerns the fluid motion inside each star.  We
consider two limiting cases: {\bf synchronized} (also called 
corotating) motion and {\bf irrotational} flow  (assuming that
the fluid has zero vorticity in the inertial frame). 
The latter state is more realistic.

We consider only {\bf equal-mass} binaries consisting of identical
 stars with gravitational masses $M_1=M_2=1.35\ M_{\odot}$ measured in
 infinite separation.  The main reason for choosing these particular
 masses is that five out of six observed binary radio pulsars have
 mass ratio close to unity and gravitational masses of each star $\sim
 1.3-1.4 M_{\odot}$ \cite{Lorim01,Burga03}.  In addition
 population synthesis calculations \cite{BulikGB04} have shown that a
 significant fraction of the observed binary neutron stars in
 gravitational waves will contain stars with equal masses $\sim 1.4\
 M_{\odot}$ and systems consisting of a low and a high mass neutron
 star.

\section{The equation of state and stellar models} 

\begin{figure}
\vskip 0.8cm
\includegraphics[width=0.45\textwidth]{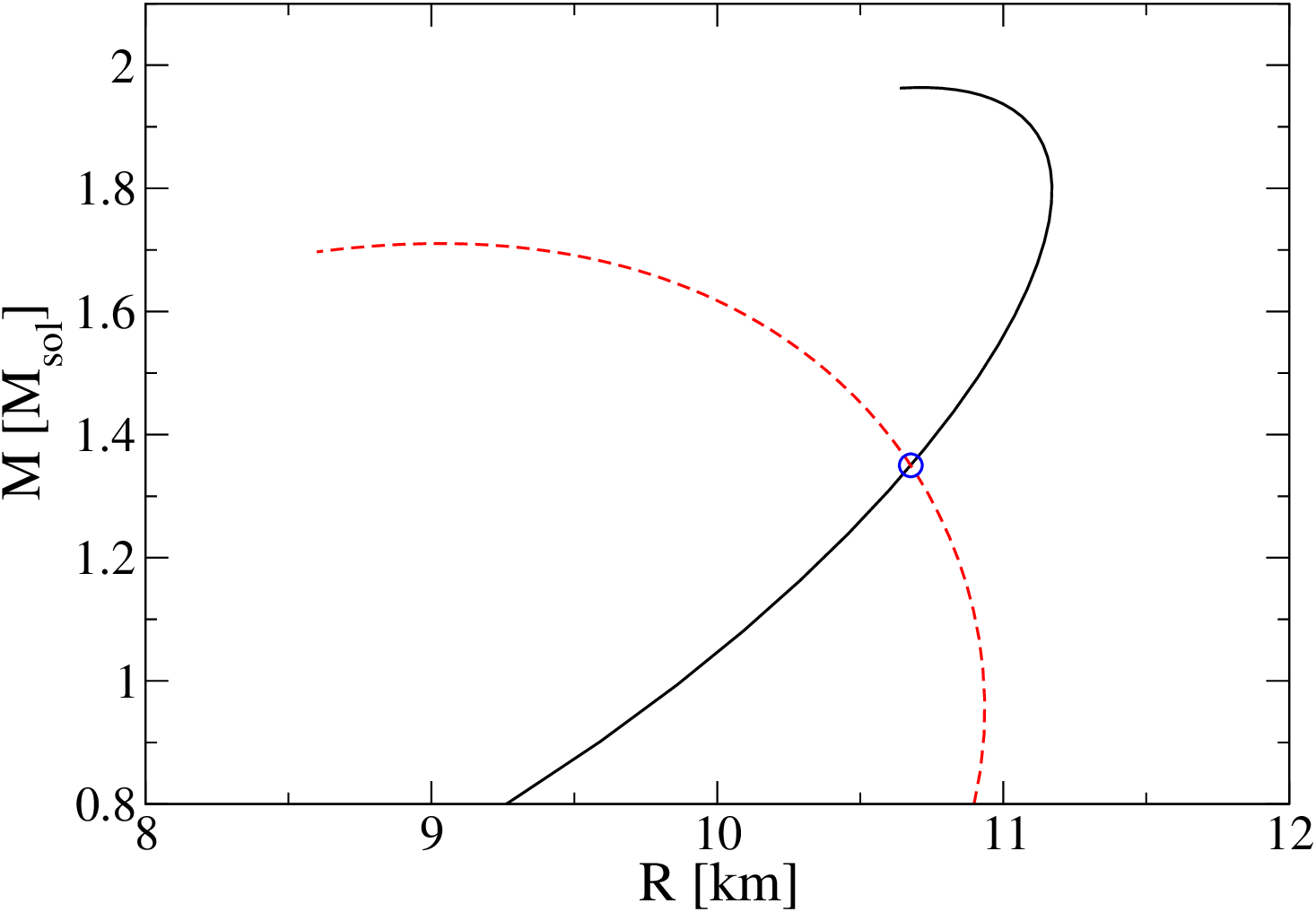}
\caption{Gravitational mass $M$ versus areal radius $R$
for sequences of
static strange quark stars described by the simplest MIT bag model 
(solid line) and neutron stars described by polytropic EOS with 
$\gamma =2.5$  and  $\kappa=0.0093\ m_0 \, n_{\rm nuc}^{1.5}$ 
(dashed line). The two sequences 
are crossing at the point $M=1.35\ M_\odot$
and $R=10.677 {\ \rm km}$ (marked by a circle). }\label{f:MRSSPOLI}
\end{figure}

It has been shown \cite{Faber02,TanigG03}
that the evolution of equal-mass binary neutron
stars depend mainly on the compactness parameter $M/R$, 
where $M$ and $R$ are the gravitational mass
measured by an observer at infinity for a single isolated neutron star
and the stellar radius respectively. It is therefore interesting to
check if the properties of inspiraling strange stars can be predicted
by studying binaries consisting of polytropic neutron stars having the
same mass and the same compactness parameter.  Therefore we perform
calculations for two different equations of state of dense matter:
a strange quark matter EOS and a polytropic EOS.

Typically, strange stars are modeled \cite{AlcocFO86, HaensZS86} with an 
equation of state based on the MIT-bag model of
quark matter, in which quark confinement is described by an energy
term proportional to the volume \cite{FahriJ84}.  The equation
of state is given by the simple formula
\beq \label{e:sseos}
p=a(\rho-\rho_0),
\eeq
\beq \label{sseosn} n(p) = n_0\cdot\left[1+{1+a\over
a}{p\over\rho_0}\right]^{1/(1+a)}, \eeq 
where $n$ is the baryon
density and $a,\ \rho_0 , n_0$ are some constants depending on the 3 parameters of the model
(the bag constant $B$, the mass of the strange quarks $m_{\rm s}$ and the
strenght of the QCD coupling constant $\alpha$). In general this
equation corresponds to a self-bound matter with mass density $\rho_0$
and baryon density $n_0$ at zero pressure and with a fixed sound
velocity ($\sqrt{a}$) at all pressures.  It was shown that different
strange quark models can be approximated very well by Eqs
(\ref{e:sseos}) and (\ref{sseosn}) \cite{Gondek00, Zdunik00}.

In the numerical calculations reported in the present paper we
describe strange quark matter using the simplest MIT bag model (with
massless and non-interacting quarks), for which the formula
(\ref{e:sseos}) is exact. We choose the value of the bag constant to
be $B=60\ {\rm MeV\ fm^{-3}}$.  For this model we have $a=1/3$,
$\rho_0=4.2785\times 10^{14}\ {\rm g\ cm}^{-3}$ and ${n_0}=0.28665\
{\rm fm^{-3}}$. In general for the MIT bag model the density of
strange quark matter at zero pressure is in the range $\sim 3 \times
10^{14}-6.5 \times 10^{14}\ {\rm g\ cm^{-3}}$ and $a$ between 0.289
and 1/3 (for $0 \le \alpha \le 0.6 \  {\rm and}\ 0 \le m_{\rm s} \le 250 \ {\rm
MeV}$) \cite{Zdunik00}. The higher value of $a$ and of
$\rho_0$ the higher compactness parameter of a star with fixed
gravitational mass.

Up to now, the majority of calculations of the hydrodynamical inspiral
phase \cite{BonazGM99,MarroMW99,UryuE00,UryuSE00,GourGTMB01,Faber02, 
TanigG02b,TanigG03,FaberGR04,MarronDSB04}
and all calculations of the merger phase 
\cite{ShibaU00,ShibaU01,ShibaTU03,OoharN99}
have been performed for
binary systems containing neutron stars described by a polytropic
EOS:
\beq
  p=\kappa n^{\gamma},
\label{eeospolyp}
\eeq where $\kappa$ and $\gamma$ coefficients are some constant numbers:
$\kappa$ represents the overall compressibility of matter
while  $\gamma$ measures the stiffness of the EOS.
The total energy density  is related to the baryon density  by
   \beq\label{eeospolye}
     e(n) = {\kappa \over \gamma-1} n^\gamma + \mu_0 \, n \ ,
   \eeq
   where  $\mu_0$ is the chemical potential at zero pressure.

In order to compare results for strange stars with those for neutron
stars, we determine the values of $\kappa$ and $\gamma$ which yield to
the same radius for the gravitational mass $M = 1.35\, M_{\odot}$ as
that obtained for a static strange star.  It was shown \cite{Bejger04}
that the properties of inspiraling neutron stars described by
realistic EOS can be, in a good approximation, predicted by studying
binaries with assumed polytropic EOSs with $\gamma=2$ or 2.5.  For a
1.35 $M_{\odot}$ strange star we have a high value of compactness
parameter $M/R=0.1867$ so we have choosen $\gamma = 2.5$, for which we found
$\kappa=0.00937 \ m_0 \, n_{\rm nuc}^{1.5}$, with  the rest mass of  relativistic particles  $m_0:=1.66\times
10^{-27} {\rm\ kg}$ and $n_{\rm nuc} = 0.1 {\ \rm fm}^{-3}$.

\begin{figure}
\vskip0.8cm
\includegraphics[width=0.46\textwidth]{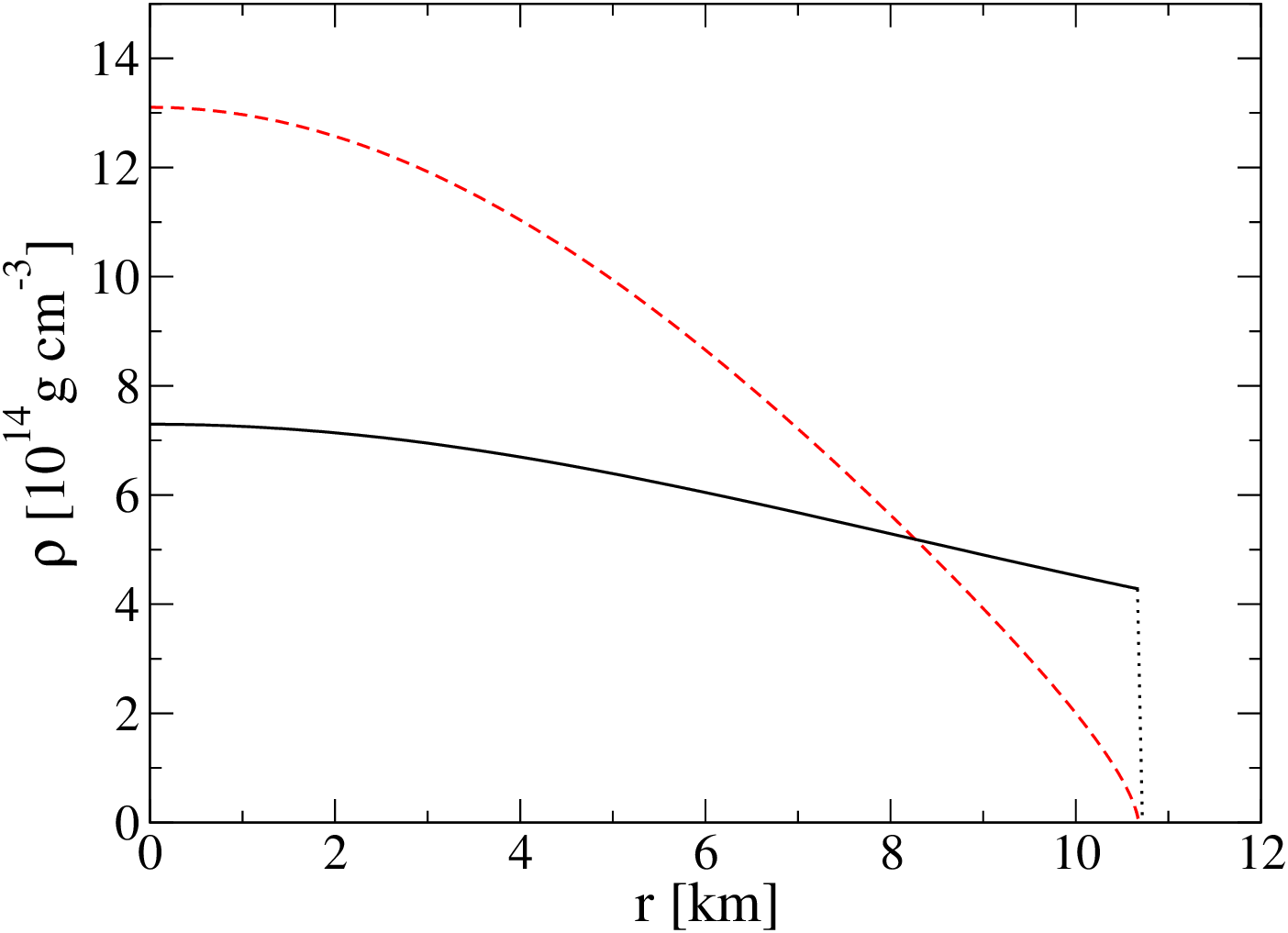}\vskip 0.9cm
\includegraphics[width=0.45\textwidth]{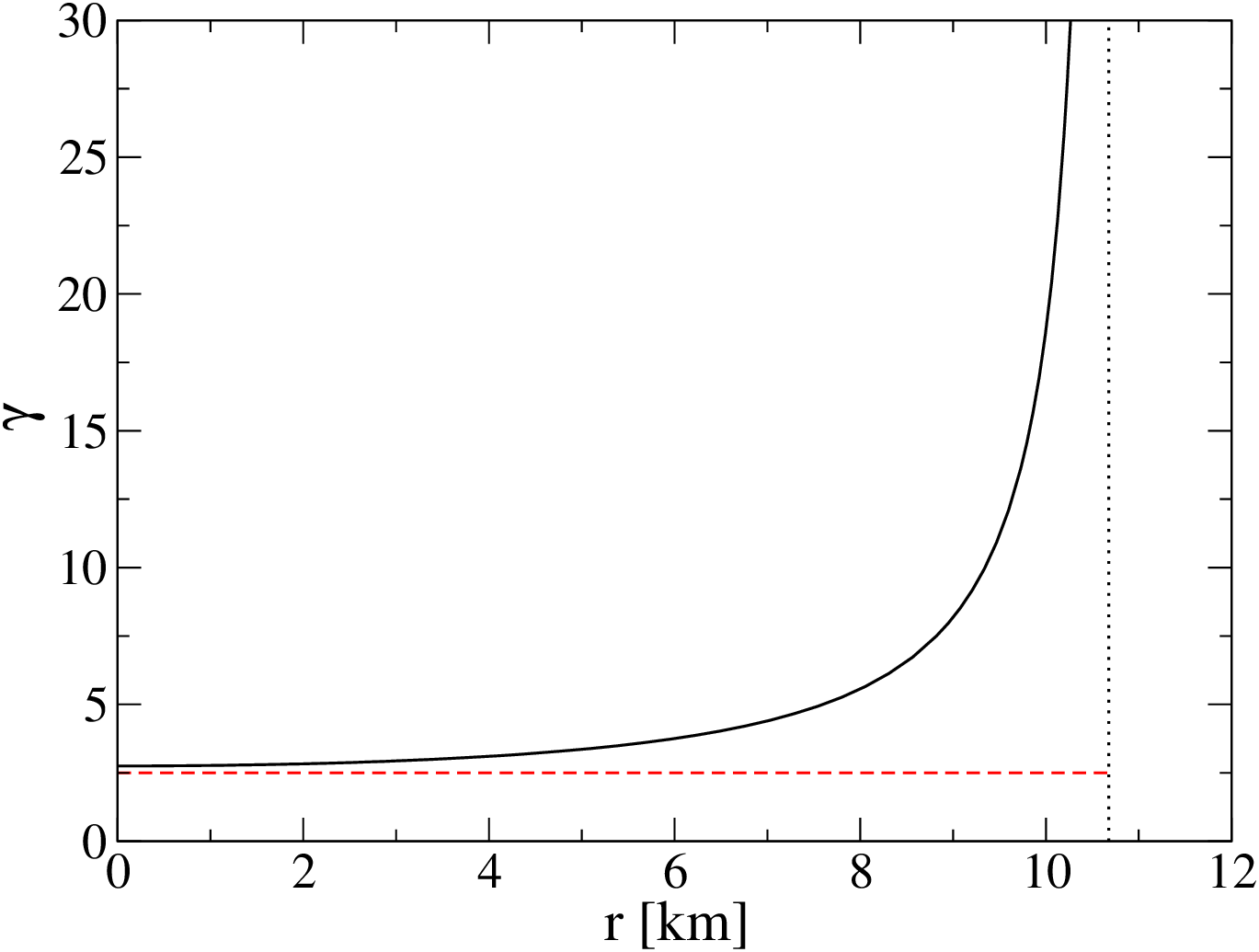}
\caption{Mass density  (top panel) and the adiabatic index  
$\gamma $ (bottom panel)  versus the  radial coordinate $r$ for a
static strange quark star (solid line) and a polytropic neutron star
(dashed line), having both a gravitational mass of $1.35\ M_\odot$ and
an areal radius $R=10.677 {\ \rm km}$ (resulting in the compactness
parameter $M/R = 0.1867$).  The vertical dotted line corresponds to
the stellar surface.}
\label{f:rho_R}
\end{figure}

In Fig. \ref{f:MRSSPOLI} we present the mass-radius relation for a
sequence of static stars described by the simplest MIT bag model
(solid line) and the polytropic EOS (dashed line) parametrized by
central density.  For small mass strange stars $M\sim R^3$ since
density is almost constant inside a star $\sim \rho_0$.
In the top panel of Fig. \ref{f:rho_R} we show the mass 
density distribution inside the
strange star (solid line) and the neutron star described by polytropic
EOS (dashed line) having gravitational mass 1.35 $M_\odot$ and areal
radius $10.667 {\rm \ km}$ 
(the configurations corresponding to the crossing point on
Fig. \ref{f:MRSSPOLI}). The huge density jump at the surface of the strange 
star corresponds to $\rho_0 = 4B$. The value of density at the surface
describes strongly or weakly bound strange matter, which in each case
must be absolutely stable with respect to ${}^{56}{\rm Fe}$.

An important quantity relevant for evolution of binary compact stars
is the adiabatic index: \beq \gamma= {\rm d}\ln p/{\rm d}\ln n .  \eeq
We assume that matter is catalized so the adiabatic index can be
calculated directly from EOS (see Refs.~\cite{GourgHG95} and
\cite{GondeHZ97} for discussion on different kind of adiabatic indices
and corresponding timescales).  Note that for the polytropic EOS given
by Eq.~(\ref{eeospolyp}) the index $\gamma$ coincides with the
adiabatic index of a relativistic isentropic fluid.  Dependence of the
adiabatic index $\gamma$ on stellar radii for both EOS is shown in the
bottom pannel of Fig.~\ref{f:rho_R}.  The adiabatic index of strange
matter is qualitatively different from the adiabatic index for
polytropic EOS or for realistic EOS.  The values of $\gamma$ in the
outer layers of strange stars are very large and for $\rho\rightarrow
\rho_0$ we have $\gamma=a+\rho/(\rho-\rho_0)\rightarrow \infty $.  The
EOS of neutron stars for densities lower than $\sim 10^{14} \ {\rm g\,
cm}^{-3}$ (the crust) is well established \cite{Haens03}.  In the
outer crust of an ordinary neutron star the pressure is dominated by
the ultra-relativistic electron gas, so we have $\gamma=4/3$.  The
values of the local adiabatic index in the inner crust of a neutron
star depends strongly on density and varies from $\gamma\simeq 0.5$
near the neutron drip point to $\gamma\simeq 1.6$ in the bottom layers
near the crust-core interface.

In our calculations we use equation of state in the form:
\beq
n=n(H),\ \ \ \ \ e=e(H),\ \ \ \ \ p=p(H), 
\eeq where H is
pseudo-enthalpy (the log-enthalpy) defined by: 
\beq \label{eeospolyh}
H(n) := \ln \left( {e+p \over n E_0}\right), 
\eeq
where the energy per unit baryon number is $E_0=m_0$ for a polytropic EOS, 
and $E_0=\rho_0 /n_0= 837.26 {\ \rm MeV}$ for strange quark model described 
above.  
For our model of strange quark matter we have:
\beq
\rho =\rho_0(3{\rm e}^{4H} + 1)/4, \ \ \ p = \rho_0({\rm e}^{4H} -1)/4~,\ \ \ \  n = n_0{\rm e}^{3H}~
\eeq

\section{Equations to be solved}

We refer the reader to Ref.~\cite{GourgGTMB01} for the derivation 
of the equations describing quasi-equilibrium binary stars within 
the IWM approximation to general relativity.
After recalling these equations, we mainly
concentrate on the equation for the velocity potential of
irrotational flows. Actually this equation has a different structure
for strange stars than for neutron stars. This results from the
non-vanishing of the density at the stellar surfaces of strange
stars (cf. the top panel in Fig.~\ref{f:rho_R}).

\subsection{The gravitational field equations}

The gravitational field equations have been obtained within the 3+1 
decomposition of the Einstein's equations \cite{York79,Cook00}, taking 
into account the helical symmetry of spacetime. 
The trace of the spatial part of the Einstein equations
combined with the Hamiltonian constraint results in two equations:
\beqa
  \ulap \nu &=& 4\pi A^2 (E+S) + A^2 K_{ij} K^{ij}
  - \unab_i \nu \unab^i \beta, \\
  \ulap \beta &=& 4\pi A^2 S +{3 \over 4} A^2 K_{ij} K^{ij} \nonumber\\
 && -{1 \over 2} (\unab_i \nu \unab^i \nu +\unab_i \beta \unab^i \beta),
\eeqa
where $\unab_i$ stands for the covariant derivative associated with
the flat 3-metric $f_{ij}$ and  $\ulap := \unab^i \unab_i$ for the 
associated Laplacian operator.
The quantities $\nu$ and $\beta$ are defined by
$\nu := \ln N$ and $\beta := \ln (AN)$, and
$K_{ij}$ denotes the extrinsic curvature tensor of the $t={\rm const}$
hypersurfaces.
$E$ and $S$ are respectively the matter energy density and the trace of the 
stress tensor, both as measured by the observer whose 4-velocity is $n^\mu$ 
{\em (Eulerian observer)}:
\beqa
  E &:=& T_{\mu \nu} n^{\mu} n^{\nu} , \\
  S &:=& A^2 f^{ij} T_{ij}.
\eeqa
In addition, we have also to solve the momentum constraint, which writes
\beqa
  \ulap N^i +{1 \over 3} \unab^i (\unab_j N^j) &=& -16\pi N A^2 (E+p) U^i \nonumber\\
   && +2 N A^2 K^{ij} \unab_j (3\beta -4\nu),
\eeqa
where $N^i:=B^i +\Omega (\p/\p \vp)^i$ denotes the shift vector of
nonrotating coordinates, and $U^i$ is the fluid 3-velocity.

\subsection{The fluid equations}

Apart from the gravitational field equations, we have to solve the
fluid equations.  The equations governing the quasi-equilibrium state
are the relativistic Euler equation and the equation of baryon number conservation.
Both cases of irrotational and synchronized motions admit a first
integral of the relativistic Euler equation:
\beq H + \nu - \ln \Gamma_0 + \ln \Gamma = {\rm
const.}, 
\eeq
where $\Gamma_0$ is the Lorentz factor between
the co-orbiting observer and
the Eulerian observer and $\Gamma$ is the Lorentz factor between the fluid 
and the co-orbiting observers ($\Gamma=1$ for synchronized binaries).

For a synchronized
motion, the equation of baryon number conservation is trivially
satisfied, whereas for an irrotational flow, it is written as 
\beqa
\label{eq:vel_pot} \zeta H \ulap \Psi &+& \unab^i H \unab_i \Psi \nonumber\\
   &=& A^2 h \Gamma_{\rm n} U^i_0 \unab_i H + \zeta H \nonumber\\
   && \times [\unab^i \Psi \unab_i
(H-\beta) +A^2 h U^i_0 \unab_i \Gamma_{\rm n}], 
\eeqa

where $\Psi$ is the velocity potential, $h := \exp(H)$,  $\zeta$ the 
thermodynamical coefficient:
\beq \label{e:zeta_h}
\zeta := {d \ln H \over d \ln n},
\eeq 
and
$\Gamma_{\rm n}$ denotes the Lorentz factor between the fluid and the 
Eulerian observer
and $U^i_0$ is the orbital 3-velocity with respect to the Eulerian
observers: \beq U^i_0 = -{B^i \over N}. \label{eq:ov_wrt_euler} \eeq

The fluid 3-velocity $U^i$ with respect to the Eulerian observer is equal to
$U^i_0$ for synchronized binary systems, whereas 
\beq \label{e:u-euler}
  U^i = {1 \over A^2 \Gamma_{\rm n} h} \unab^i \Psi \label{eq:iv_wrt_corot}
\eeq
for irrotational ones.

\subsection{Boundary condition for the velocity potential}

The method of solving the elliptic equation (\ref{eq:vel_pot}) for the
velocity potential is different for neutron stars and strange stars.
In the case of neutron stars, the coefficient $\zeta H$ in front of
the Laplacian vanishes at the surface of the star so
Eq. (\ref{eq:vel_pot}) is not merely a Poisson type equation for
$\Psi$.  It therefore deserves a special treatment (see Appendix B in
\cite{GourGTMB01} for a discussion).  In the case of strange stars,
the coefficient $\zeta H = 1/3$ in whole star so we have to deal with
a usual Poisson equation and consequently we have to impose a boundary
condition for the velocity potential at the stellar surface. 

We can define the surface of the star by $n_{| \rm{surf}} = n_0 = 
\rm{constant}$.  
The surface of the fluid ball is obviously Lie-dragged along the fluid 
4-velocity vector $\w{u}$, so that this last condition gives
\beq \label{e:lie1}
\left.  (\pounds _{\w{u}} n) \right| _{\rm{surf}} = 0,
\eeq
where $\pounds _{\w{u}}$ is the Lie derivative along the vector field
$\w{u}$.
Let us decompose $\w{u}$ in a part along the helical Killing vector 
$\w {\ell}$ and a part $\w{S}$ parrallel to the hypersurface $\Sigma_t$:
\beq \label{e:decomposition}
\w{u} = \lambda (\w {\ell} + \w{S}) .
\eeq
The condition (\ref{e:lie1}) is then equivalent to
\beq  \label{e:lie2}
\left. \lambda ( \pounds _{\w {\ell}} n 
+ \pounds _{\w{S}} n ) \right| _{\rm{surf}} = 0.
\eeq
Now, if the fluid flows obeys to the helical symmetry
$\pounds _{\w {\ell}} n = 0$; inserting this relation 
into Eq.~(\ref{e:lie2}) leads to 
$\left. (\pounds _{\w{S}} n ) \right| _{\rm{surf}} = 0$
or equivalently (since $\w{S}$ is a spatial vector):
\beq \label{e:condition1}
\left. (S^i  \unab_i n) \right| _{\rm{surf}} = 0.
\eeq
Now, let us express $\w{S}$ in terms of the spatial vectors 
$\w{U}$ and $\w{B}$. First,  Eq. ~(\ref{e:decomposition}) implies 
$\w{n} \cdot \w{u} = \lambda \w{n} \cdot \w {\ell}$.
Secondly, the fluid motion $\w{u}$ can be described by the orthogonal 
decomposition $\w{u} = \Gamma_{\rm n} (\w{n} + \w{U})$ which yields 
$\w{n} \cdot \w{u} = - \Gamma_n$.
Finally, from Eq. ~(\ref{e:helicoidal_n}), we have 
$\w{n} \cdot \w {\ell} = - N$ so that the factor $\lambda$ can be expressed as 
$\lambda = \Gamma_n / N$
and Eq. ~(\ref{e:decomposition}) becomes 
\beq \label{e:decomposition2}
\w{u} = {\Gamma_n \over N} (\w {\ell} + \w{S}).
\eeq
Now, combining  Eq. ~(\ref{e:decomposition2}) 
and  Eq. ~(\ref{e:helicoidal_n}), we have 
\beq \label{e:decomposition3}
\w{u} = \Gamma_n \left[ \w{n} + {1 \over N} (\w{S} - \w{B}) \right] .
\eeq
Comparing with the orthogonal decomposition
$\w{u} = \Gamma_{\rm n} (\w{n} + \w{U})$, we deduce that
$\w{S} = N \w{U} + \w{B}.$
Inserting this relation into Eq.~(\ref{e:condition1}) leads to the boundary 
condition
\beq
\left. (N U^i \unab_i n + B^i \unab_i n) \right| _{\rm{surf}} = 0.
\eeq
Now, using Eq. ~(\ref{e:u-euler}), we obtain a Neumann-like boundary condition 
for $\Psi$:
\beq \label{e:condition2}
\left. (\unab^i n \unab_i \Psi) \right| _{\rm{surf}} 
= - \left. \left({\Gamma_n h A^2 \over N} B^i \unab_i n\right)
        \right| _{\rm{surf}} .
\eeq
Considering the elliptic equation (\ref{eq:vel_pot}) for $\Psi$ we 
see that the boundary condition we have obtained is consistent with the 
case $n = 0$ (or equivalently $\zeta H = 0$) at the stellar surface since, 
from Eq. ~(\ref{e:zeta_h}), $\unab^i H = {\zeta H \over n} \unab^i n$.

\section{Numerical results}

\subsection{The method}

The resolution of the above nonlinear elliptic equations is 
performed thanks to a numerical code based on multidomain spectral 
methods and constructed upon the {\sc Lorene} C++ library \cite{Lorene}. 
The detailed description of the whole algorithm, 
as well as numerous tests of the code can be found in \cite{GourgGTMB01}.
Additional tests have been presented in Sec.~3 of \cite{TanigG03}.
The code has
already been used successfully for calculating the final phase of
inspiral of binary neutron stars described by polytropic EOS
\cite{BonazGM99,TanigGB01,TanigG02a,TanigG02b,TanigG03} and realistic EOS
\cite{Bejger04}. 
It is worth to stress that the adaptation of the 
domains (numerical grids) to the stellar surface (surface-fitted coordinates)
used in this code is particulary usefull here, 
due to the strong discontinuity of the density field at the surface of strange
stars (cf. the top panel in Fig.~\ref{f:rho_R}). Adapting the grids to the stellar surface
allows to avoid the severe Gibbs phenomenon that such a discontinuity
would necessary generate when performing polynomial expansions of the
fields \cite{BonazGM98}.

The hydrodynamical part of the code has been amended for the 
present purpose, namely to solve Eq.~(\ref{eq:vel_pot}) for the
velocity potential $\Psi$ subject to the boundary condition 
(\ref{e:condition2}). Let us recall that in the original version
of the code, the treatment of Eq.~(\ref{eq:vel_pot}) was different
due to the vanishing of the density field at the stellar surface
(see Appendix~B of Ref.~\cite{GourgGTMB01}). 

We have used one numerical domain for each star and 
3 (resp. 4) domains for the space around
them for a small (resp. large) separation. In each domain, the number
of collocation points of the spectral method is chosen to be $N_r
\times N_{\theta} \times N_{\varphi} = 25 \times 17 \times 16$, where
$N_r$, $N_{\theta}$, and $N_{\varphi}$ denote the number of 
collocation points ($=$ number of polynomials used in the spectral
method) in
the radial, polar, and azimuthal directions respectively.  The
accuracy of the computed relativistic models has been estimated using a
relativistic generalization of the virial theorem
\cite{FriedUS02} (see also Sec.~III.A of Ref.  \cite{TanigG03}). The
virial relative error is a few times $10^{-5}$ for the closest
configurations.

\subsection{Evolutionary sequences}

For each EOS we construct an {\em evolutionary sequence}, i.e. a
sequence of quasi-equilibrium configurations 
with fixed baryon mass and decreasing separation.  Such a sequence is
expected to approximate pretty well the true evolution of binary
neutron stars, which is entirely driven by the reaction to
gravitational radiation and hence occur at fixed baryon number and
fluid circulation. 

\begin{figure}{}
\vskip 0.5cm
\includegraphics[angle=0,width=0.45\textwidth]{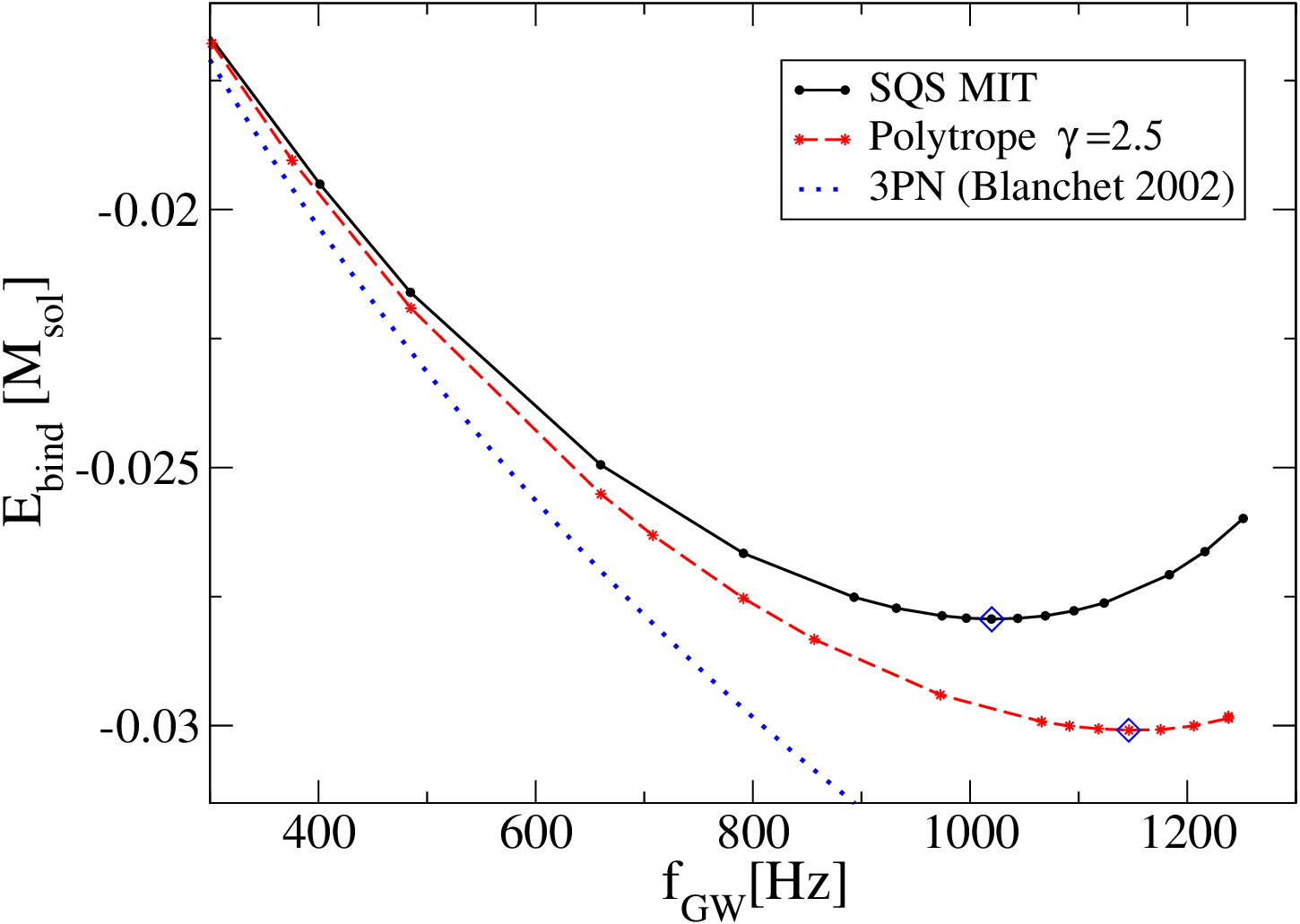}
\caption{ Binding energy as a function of gravitational wave frequency
 along evolutionary sequences of corotating binaries.  The solid line
 denotes strange quark stars, the dashed one neutron stars with
 polytropic EOS, and the dotted one point-mass binaries in the 3PN
 approximation \cite{Blanc02}. The diamonds locate the minimum of the
 curves, corresponding to the innermost stable circular orbit;
 configurations to the right of the diamond are securaly
 unstable. }\label{f:EJ_cor}
\end{figure}

For a given rotational state we calculate evolutionary sequences of
binary system composed of two identical neutron stars or two identical
strange stars. The evolution of inspiraling corotating (irrotational)
binaries is shown in Fig. \ref{f:EJ_cor} (Fig. \ref{f:EJ_irr}). 
Fig. \ref{f:EJ_cor} and upper panel of Fig. \ref{f:EJ_irr} 
show the binding energy $E_{\rm bind}$ versus frequency of
gravitational waves $f_{\rm GW}$ and lower panel of Fig. \ref{f:EJ_irr}
show the total
angular momentum of the systems as a function of $f_{\rm GW}$.  The
binding energy is defined as the difference between the actual ADM
mass of the system, $M_{\rm ADM}$, and the ADM mass at infinite
separation ($2.7 \, M_{\odot}$ in our case). The frequency of
gravitational waves is twice the orbital frequency, since it
corresponds to the frequency of the dominant part $l=2,\; m=\pm 2$.
Solid and dashed lines denote quasi-equilibrium sequences of strange
quark stars binaries and neutron stars binaries respectively. Dotted
lines in Fig. \ref{f:EJ_cor} and Fig. \ref{f:EJ_irr} correspond to the
3rd PN approximation for point masses derived by \cite{Blanc02}.
Finally in Fig. \ref{f:Ebind_f_all} we compare our results with third
order post-Newtonian results for point-mass particles obtained in the
effective one body approach by Damour et al. 2000 \cite{DamourJS},
Damour et al. 2002 \cite{DamouGG02} and in the standard nonresummed
post-Newtonian framework by Blanchet 2002 \cite{Blanc02}.

A turning point of $E_{\rm bind}$ along an evolutionary sequence
indicates an orbital instability \cite{FriedUS02}.
This instability originates both from relativistic effects (the well-known
$r=6M$ last stable orbit of Schwarzschild metric) and hydrodynamical
effects (for instance, such an instability exists for sufficiently
stiff EOS in the Newtonian regime, see e.g. \cite{TanigGB01} and 
references therein). It is secular for
synchronized systems and dynamical for irrotational ones.

In the case where no turning point of $E_{\rm bind}$ occurs along the
sequence, the mass-shedding limit (Roche lobe overflow) marks the
end of the inspiral phase of the binary system, since recent dynamical
calculations for $\gamma = 2$ polytrope have shown that the time to
coalescence was shorter than one orbital period for configurations at
the mass-shedding limit \cite{ShibaU01, MarronDSB04}.  Thus the
physical inspiral of binary compact stars terminates by either the
orbital instability (turning point of $E_{\rm bind}$) or the
mass-shedding limit. In both cases, this defines the {\em innermost
stable circular orbit (ISCO)}.  The orbital frequency at the ISCO is a
potentially observable parameter by the gravitational wave detectors,
and thus a very interesting quantity.

\subsection{Corotating binaries}

Quasi-equilibrium sequences of equal mass corotating binary neutron
stars and strange stars are presented in Fig. \ref{f:EJ_cor}.  For
both sequences we find a minimum of the binding energy. 
In the present rotation state, this locates
a secular instability \cite{FriedUS02}.  
The important difference between neutron stars and
strange stars is the frequency at which this
instability appears. Indeed, there is a difference of more than 100~Hz: 
1020~Hz for strange stars and 1140~Hz for neutron stars. The binding
energy is the total energy of gravitational waves emitted by the
system: a corotating binary strange star system emits less energy 
in gravitational
waves and loses less angular momentum before the ISCO than a binary neutron
star one with the same mass and compaction parameter in infinite
separation.

Comparison of our numerical results with 3rd order PN
 calculations reveals a good agreement for small frequencies (large
 separations) (see Fig. \ref{f:EJ_cor} and
 \ref{f:Ebind_f_all}).  The deviation from PN curves 
 at higher frequencies (smaller separation) is due to hydrodynamical
 effects, which are not taken into account in the PN approach.

\subsection{Irrotational binaries}

\begin{figure}{}
\vskip 0.5cm
\includegraphics[angle=0,width=0.45\textwidth]{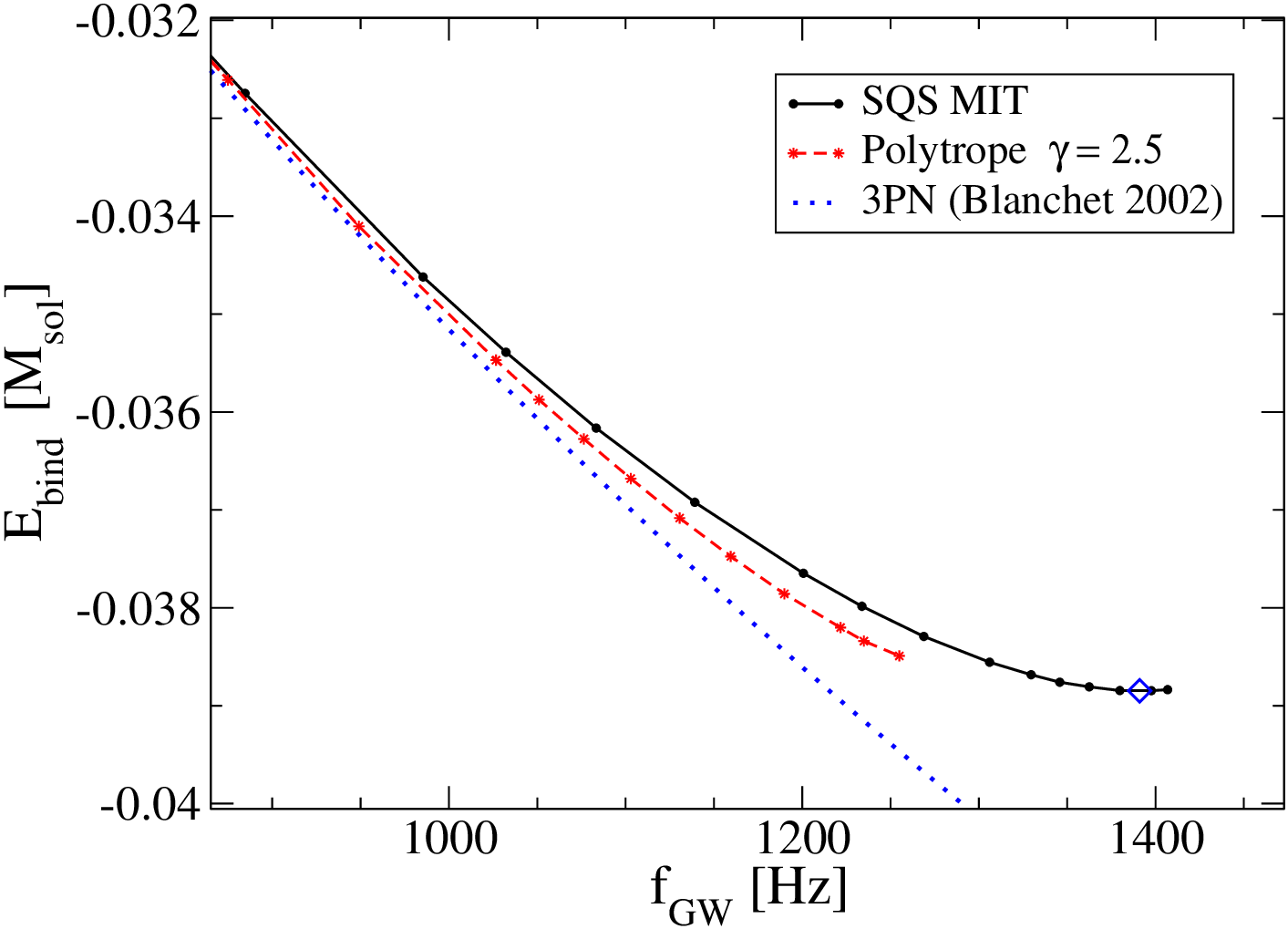} \qquad \\
\vskip 1.2cm
\includegraphics[angle=0,width=0.45\textwidth]{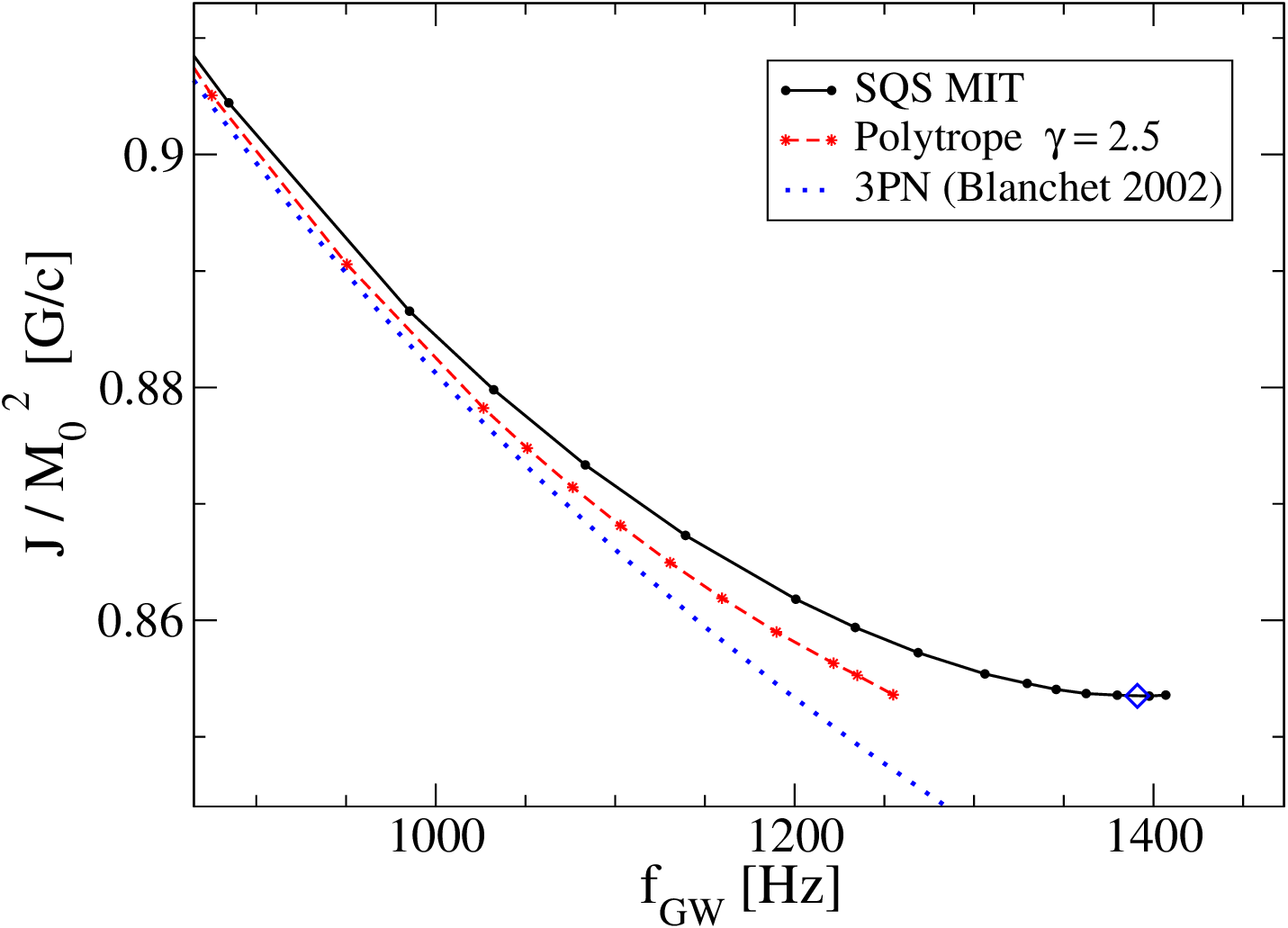}
\caption{Binding energy (top panel) and angular momentum (bottom
panel) as a function of gravitational wave frequency  along
evolutionary sequences of irrotational binaries. The solid line denotes
strange quark stars, the dashed one polytropic neutron stars, 
and the dotted one point-mass binaries in the 3PN approximation \cite{Blanc02}.
The diamonds correspond to dynamical
orbital instability (the ISCO).}
\label{f:EJ_irr}
\end{figure}

\begin{figure}{}
\vskip 0.5cm
\includegraphics[angle=-90,width=0.35\textwidth]{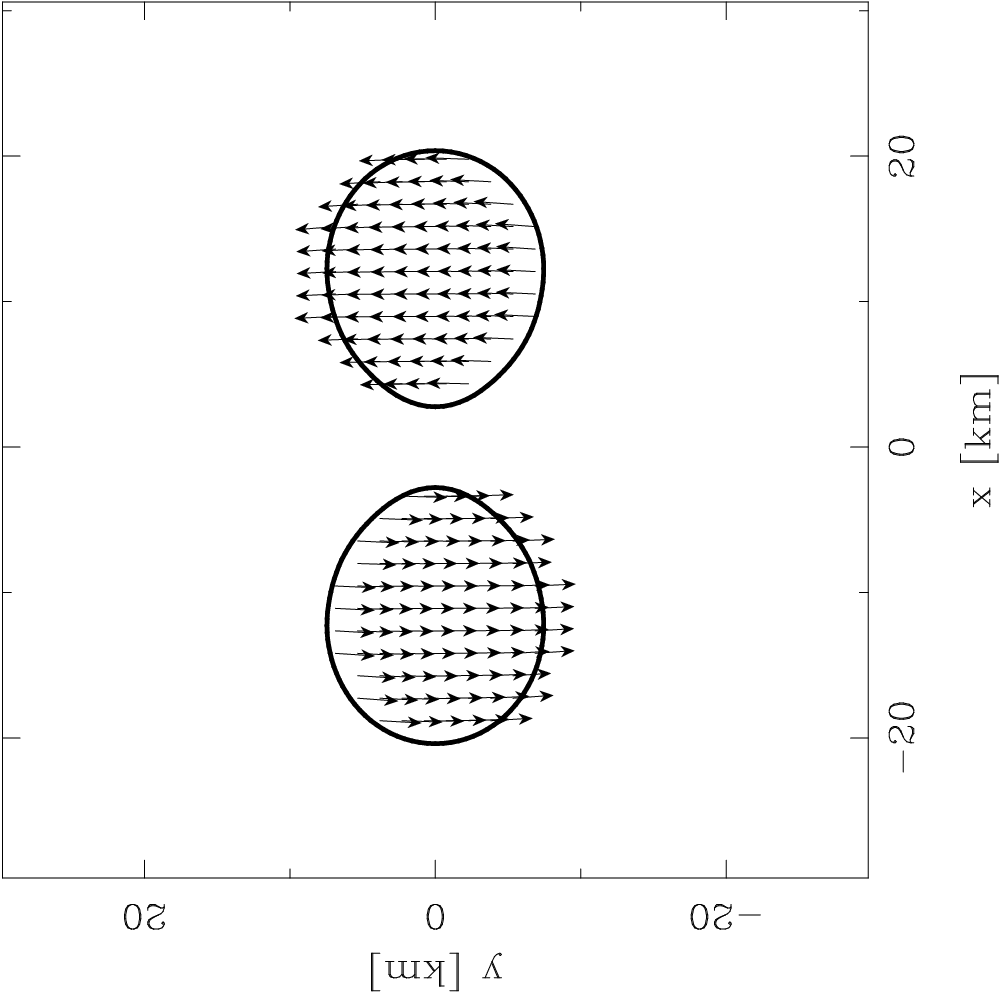}\qquad
\vskip 0.5cm
\includegraphics[angle=-90,width=0.35\textwidth]{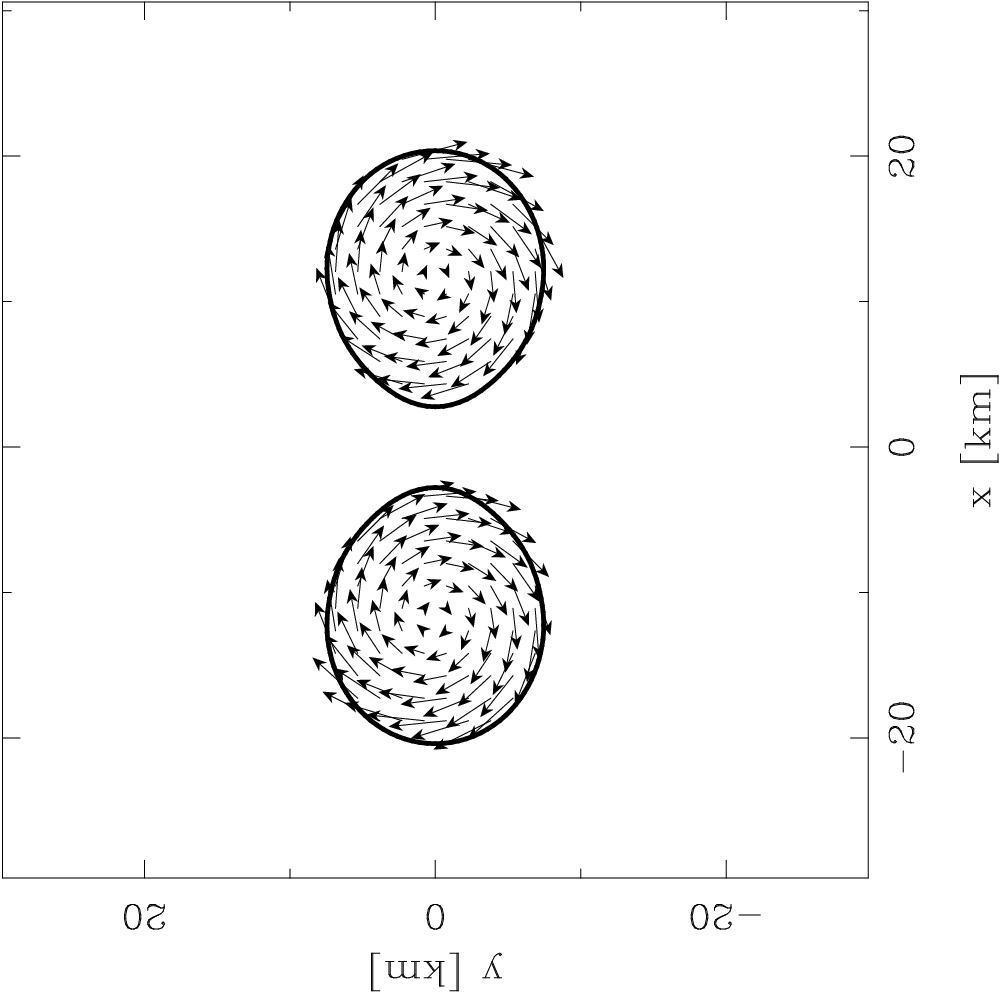}
\caption{Internal velocity fields of irrotational strange quark stars binaries
at the ISCO. {\em upper panel:} velocity $\w{U}$ in the orbital plane 
with respect to the ``inertial'' frame (Eulerian observer);
{\em lower panel:} velocity field with respect to the
corotating frame.  
The thick solid lines denote the surface of each star. }\label{f:psi0}
\end{figure}

\begin{figure}{}
\vskip 1cm
\includegraphics[angle=0,width=0.45\textwidth]{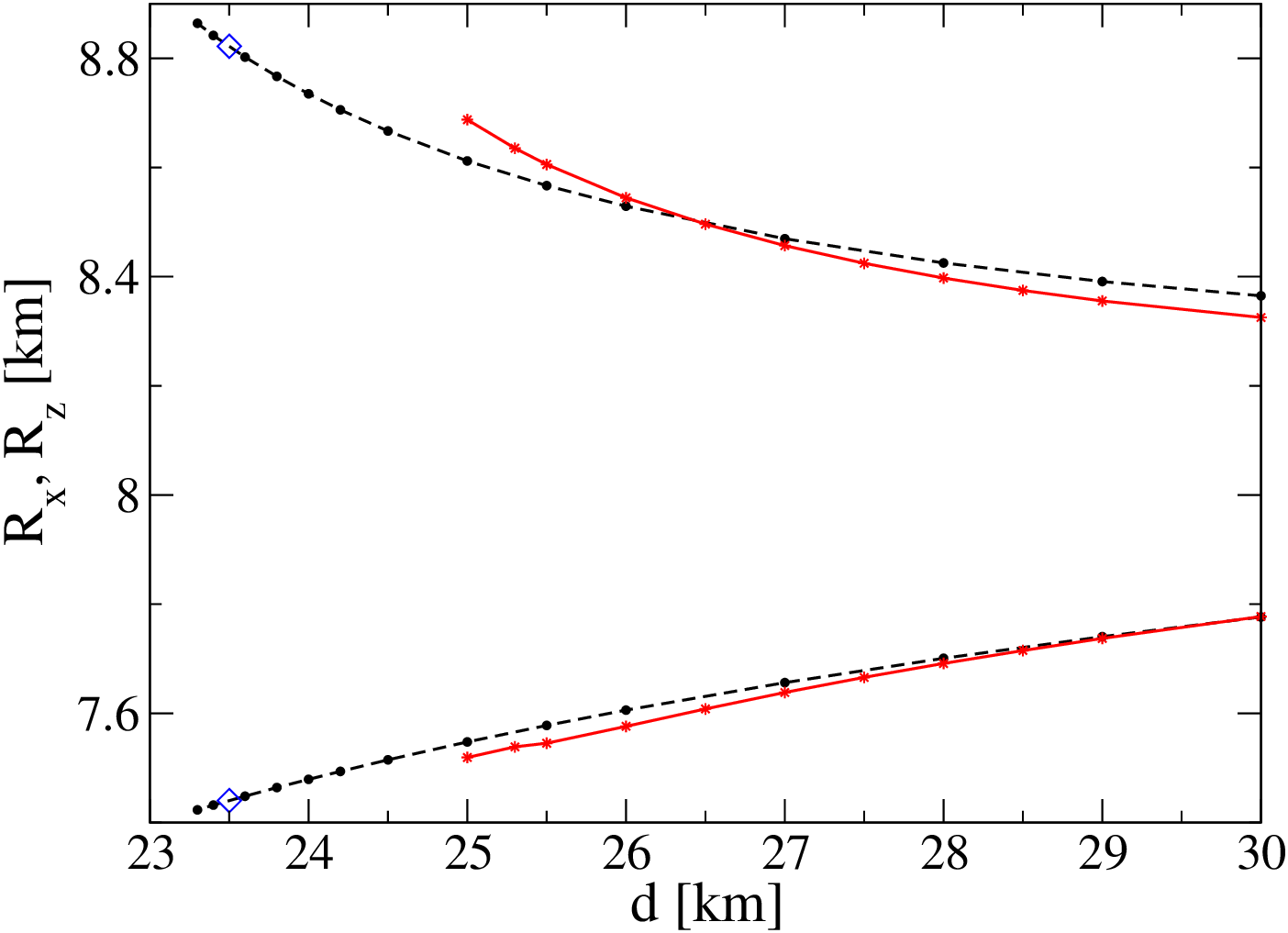}

\caption{Coordinate ``radius'' (half the coordinate size of a star in
fixed direction) versus coordinate separation for
irrotational quasi-equilibrium sequences of binary strange stars
(solid line) and neutron stars (dashed line). Upper lines correspond
to equatorial radius $R_x$ (radius along the x axis going through
the centers of stars in a binary system) and lower lines are polar
radius $R_z$ (radius along the rotation axis).}
\label{f:a1a2}
\end{figure}

In Fig. \ref{f:EJ_irr} we present the evolution of the 
binding energy and angular momentum
for irrotational sequences of binary neutron stars and strange stars. 
We also verify that these sequences are in a good agreement with PN
calculations for large separations.

We note important differences in the
evolution of binary systems consisting of strange stars or
neutron stars.
The strange star sequence shows a minimum of the
binding energy at $f_{\rm GW}\simeq 1390{\rm\ Hz}$, 
which locates a dynamical instability \cite{FriedUS02}
and thus defines the ISCO.  
The minimum of $E_{\rm bind}$ coincides with the minimum of 
total angular momentum $J$. This is in accordance with the ``first
law of binary relativistic star thermodynamics'' 
within the IWM approximation
as derived by Friedman, Uryu and Shibata \cite{FriedUS02}
and which states that, along an evolutionary sequence,
\beq
    \delta M_{\rm ADM} = \Omega \delta J.
\eeq
The surface of strange stars at the ISCO is smooth 
(see Fig. \ref{f:psi0}).
On the contrary the neutron star sequence does not present any
turning point of $E_{\rm bind}$, so that the ISCO 
in this case corresponds to the mass-shedding limit
(final point on the dashed curves in Fig.~\ref{f:EJ_irr}). 
The gravitational wave frequency at the ISCO 
is much lower for neutron star binaries than for strange star binaries. 

As already mentioned the adiabatic index in the outer layers of a
compact star in a binary system plays a crucial role in its evolution,
especially in setting the mass-shedding limit.  Although the crust of
a $1.35\, M_\odot$ neutron star contains only a few percent of the
stellar mass, this region is easily deformed under the action of the
tidal forces resulting from the gravitational field produced by the
companion star. The end of inspiral phase of binary stars strongly
depends on the stiffness of matter in this region.  It has been shown
that the turning-point orbital instability for irrotational polytropic
neutron stars binaries can be found only if $\gamma \ge 2.5$ and if
the compaction parameter is smaller than certain value
(\cite{TanigG03}, \cite{UryuSE00}).  In fact, as shown in Fig. 31 of
paper \cite{TanigG03}, they didn't find ISCO for irrotational binary
neutron stars with $\gamma = 2.5$ or $\gamma = 3$ for compaction
parameter as high as $M/R = 0.187$.

 In Fig.~\ref{f:a1a2} we present the evolution of two different
 stellar radii: the {\em equatorial radius} $R_x$, defined as half the
 diameter in the direction of the companion and the {\em polar
 radius}, defined as half the diameter parallel to the rotation axis.
 For spherical stars $R_x=R_z$. We see that at the end of the inspiral
 phase, neutron stars are, for the same separation, more oblate (more
 deformed) than strange stars. Binary neutron stars reach the
 mass-shedding limit (the point at which they start to exchange matter -
 a cusp form at the stellar surface in the direction of the
 companion) at coordinate separation $d\sim 25 {\rm km}$. We don't see
 any cusps for strange stars even for distances slightly smaller that
 the distance corresponding to the ISCO $\sim 23.5\ {\rm km}$.

 It is worth to remind here the results on
 rapidly rotating strange stars and neutron stars.  The Keplerian
 limit is obtained for higher oblatness (more deformed stars),
 measured for example by the ratio of polar and equatorial radius, in
 the case of strange stars than in the case of neutron stars
 \cite{Cook94, Sterg99,Gondek00, ZduniHGG00, AmsteBGK00, Gondek01}.

The differences in the evolution of binary (or rotating)
strange stars and neutron stars stem from the fact that strange stars
are principally bound by another force than gravitation: the
strong interaction between quarks.

\begin{figure*}{}
\vskip 0.5cm
\includegraphics[angle=0,width=0.75\textwidth ]{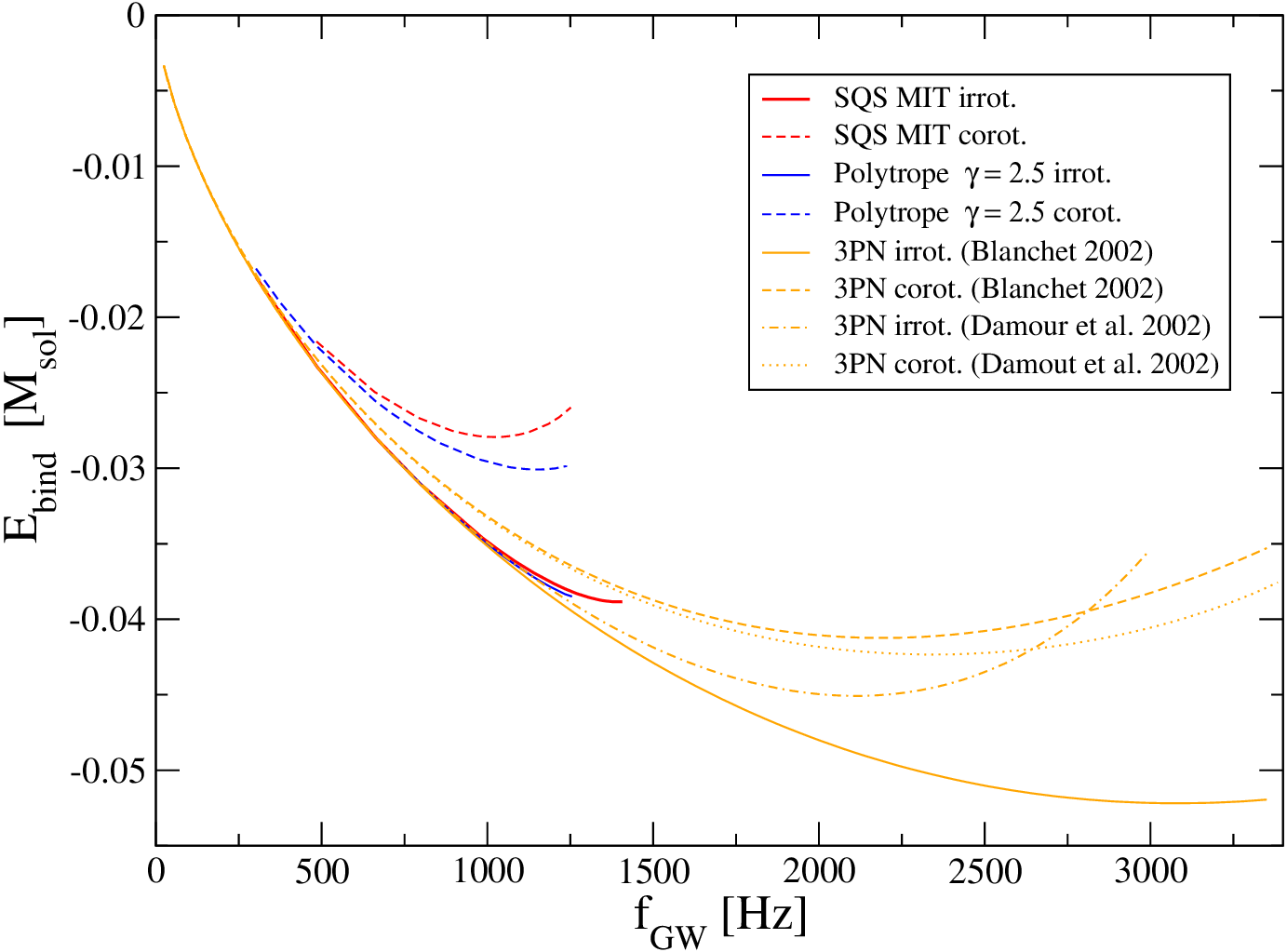} \qquad
\caption{Binding energy versus frequency of gravitational waves along
evolutionary sequences of corotational (thick dashed lines) and
irrotational (thick solid lines) equal mass (of $1.35\ M_{\odot}$)
strange stars and polytropic neutron stars binaries compared with
analytical results at the 3rd post-Newtonian order for point-masses by
Damour et al. 2000 \cite{DamourJS}, Damour et al. 2002
\cite{DamouGG02} and Blanchet 2002 \cite{Blanc02} }
\label{f:Ebind_f_all}
\end{figure*}

As already mentioned the frequency of gravitational waves is one of
potentially observable parameters by the gravitational wave detectors.
We can see from Fig. \ref{f:Ebind_f_all} that the 3rd PN
approximations for point masses derived by different authors are
giving ISCO at very high frequencies of gravitational waves $> 2\ {\rm
kHz}$. Since in the hydrodynamical phase of inspiral the effect of a
finite size of the star (e.g.  tidal forces) is very important we see
deviation of our numerical results from point-masses calculations. The
frequency of gravitational waves at the ISCO strongly depends on
equation of state and the rotational state. For irrotational equal
mass (of 1.35 $M_{\odot}$ at infinite separation) strange stars
binaries described by the simple MIT bag model this frequency is $\sim
1400 {\rm Hz}$ and for neutron stars binaries described by four
different realistic EOS it is between $ 800{\rm Hz}$ and $1230{\rm
Hz}$ \cite{Oechslin04, Bejger04}.

\section{Summary and discussion}
We have computed evolutionary sequences of irrotational and corotating
 binary strange stars by keeping the baryon mass constant to a value
 that corresponds to individual gravitational masses of $1.35\,
 M_\odot$ at infinite separation. The last orbits of inspiraling
 binary strange stars have been studied in the quasi-equilibrium
 approximation and in the framework of Isenberg-Wilson-Mathews
 approximation of general relativity. In order to calculate
 hydrodynamical phase of inspiraling irrotational strange stars
 binaries, i.e. assuming that the fluid has zero vorticity in the
 inertial frame, we found the boundary condition for the velocity
 potential. This boundary condition is valid for both the case of
 non-vanishing (e.g. self-bound matter) and vanishing density at the
 stellar surface (neutron star matter). In our calculations strange
 stars are built by strange quark matter described by
 the simplest MIT bag model (assuming massless and non-interacting
 quarks).

We have located the end of each quasi-equilibrium sequence (ISCO),
 which corresponds to some orbital instabilities (the dynamical
 instability for irrotational case or the secular one for synchronized
 case) and determined the frequency of gravitational waves at this
 point.  This characteristic frequency yields important information
 about the equation of state of compact stars and is one of the
 potentially observable parameters by the gravitational wave
 detectors.  In addition, the obtained configurations provide valuable
 initial conditions for the merger phase. We found the frequency of
 gravitational waves at the ISCO to be $\sim 1400\ {\rm Hz}$ for
 irrotational strange star binaries and $\sim 1000\ {\rm Hz}$ for
 synchronized case.  The irrotational case is more realistic since the
 viscosity of strange star matter is far too low to ensure
 synchronization during the late stage of the inspiral.  For
 irrotational equal mass (of 1.35 $M_{\odot}$) neutron star binaries
 described by realistic EOS \cite{Oechslin04, Bejger04} the frequency
 of gravitational waves at the ISCO is between $ 800{\rm Hz}$ and
 $1230{\rm Hz}$, much lower than for a binary strange quark star built
 of self-bound strange quark matter. We have considered only strange
 quark stars described by the simple MIT bag model with massless and
 non-interacting quarks.  In order to be able to interpret future
 gravitational-wave observations correctly it is necessary to perform
 calculations for different strange star EOS parameters (taking also
 into account the existence of a thin crust) and for large sample of
 neutron stars described by realistic equations of state. For some MIT
 bag model parameters one is able to obtain less compact stars than
 considered in the present paper. In this case the frequency of
 gravitational waves at the end of inspiral phase will be lower than
 obtained by us. It should be also taken into account that stars in a
 binary system can have different masses \cite{BulikGB04}.  The case
 of binary stars (with equal masses and different masses) described by
 different strange quark matter models will be presented in a separate
 paper \cite{GondeL05}.
 
 We have shown the differences in the inspiral phase between strange quark
 stars and neutron stars described by polytropic equation of state
 having the same gravitational mass and radius in the infinite
 separation. It was already shown by Bejger et al. 2005
 \cite{Bejger04} that the frequency of gravitational waves at the end
 point of inspiraling neutron stars described by several realistic EOS
 without exotic phases (such as meson condensates or quark matter) can
 be predicted, in a good approximation, by studying binaries with
 assumed polytropic EOSs with $\gamma=2$ or 2.5.  For realistic EOS
 and polytropes with $\gamma \le 2.5$ \cite{UryuSE00, TanigG03} a
 quasi-equilibrium irrotational sequence terminates by mass-shedding
 limit (where a cusp on the stellar surface develops).

We found that it wasn't the case for inspiraling strange star binaries
which are self-bound objects having very large adiabatic index in the
outer layer. For both synchronized and irrotational configurations, we
could always find a turning point of binding energy along an
evolutionary sequence of strange quark stars, which defines an orbital
instability and thus marks the ISCO in this case. In the irrotational
case for the same separation strange stars are less deformed than
polytropic neutron stars and for the same ratio of coordinate radius
$R_x/R_z$ their surfaces are more smooth. A cusp doesn't appear on
the surface of a strange star in a binary system even for separation
corresponding to orbital instability.  The frequency of gravitational
waves at the end of inspiral phase is higher by 300 Hz for the strange
star binary system than for the polytropic neutron star binaries.
The differences in the evolution of binary (or rotating) strange stars
and neutron stars stem from the fact that strange stars are
principally bound by an additional force, strong interaction between
quarks.
 
\acknowledgements We thank our anonymous referee for helpful comments.
Partially supported by the KBN grants 5P03D.017.21 and
PBZ-KBN-054/P03/2001; by ``Ayudas para movilidad de Profesores de
Universidad e Invesigadores espanoles y extranjeros'' from the Spanish
MEC; by the ``Bourses de recherche 2004 de la Ville de Paris'' and by
the Associated European Laboratory Astro-PF (Astrophysics
Poland-France).


\end{document}